# Laboratory Studies for Planetary Sciences
# A Planetary Decadal Survey White Paper
# Prepared by the American Astronomical Society (AAS) Working Group on Laboratory Astrophysics (WGLA)

http://www.aas.org/labastro

*Targeted Panels: Terrestrial Planets; Outer Solar System Satellites; Small Bodies*

Lead Author:
Murthy S. Gudipati
Ice Spectroscopy Lab, Science Division, Mail Stop 183-301, Jet Propulsion Laboratory, California Institute of Technology, 4800 Oak Grove Drive, Pasadena, CA 91109.
gudipati@jpl.nasa.gov, 818-354-2637

Co-Authors:
Michael A'Hearn - University of Maryland
ma@astro.umd.edu, 301-405-6076
Nancy Brickhouse - Harvard-Smithsonian Center for Astrophysics
nbrickhouse@cfa.harvard.edu, 617-495-7438
John Cowan - University of Oklahoma
cowan@nhn.ou.edu, 405-325-3961
Paul Drake - University of Michigan
rpdrake@umich.edu, 734-763-4072
Steven Federman - University of Toledo
steven.federman@utoledo.edu, 419-530-2652
Gary Ferland - University of Kentucky
gary@pa.uky.edu, 859-257-879
Adam Frank - University of Rochester
afrank@pas.rochester.edu, 585-275-1717
Wick Haxton - University of Washington
haxton@u.washington.edu, 206-685-2397
Eric Herbst - Ohio State University
herbst@mps.ohio-state.edu, 614-292-6951
Michael Mumma - NASA/GSFC
michael.j.mumma@nasa.gov, 301-286-6994
Farid Salama - NASA/Ames Research Center
Farid.Salama@nasa.gov, 650-604-3384
Daniel Wolf Savin - Columbia University
savin@astro.columbia.edu, 212-854-4124,
Lucy Ziurys – University of Arizona
lziurys@as.arizona.edu, 520-621-6525

**Brief Description:**

The WGLA of the AAS promotes collaboration and exchange of knowledge between astronomy and planetary sciences and the laboratory sciences (physics, chemistry, and biology). Laboratory data needs of ongoing and next generation planetary science missions are carefully evaluated and recommended.

**Introduction:**

With the successful outer solar system missions Galileo and Cassini, and the in-situ rover/lander missions to the inner solar system planet Mars in the past few decades, planetary sciences is presently in its golden era. In this highly successful space science endeavor, laboratory studies have played a crucial role, along with new missions, refined models, and novel instruments. For example, detection of molecular oxygen on Ganymede (Spencer et al., 1995) was possible only due to the laboratory data that is available on the molecular oxygen dimmer $(O_2)_2$ absorption at 627.5 and 577.3 nm in solid and liquid oxygen (Landau et al., 1962). Similarly, hydrogen peroxide ($H_2O_2$) was positively detected on Europa's surface with the help of laboratory studies (Carlson et al., 1999). On the other hand, lack of laboratory data in the far infrared region below 100 cm$^{-1}$ for water ice analogs of Saturn's rings still hamper the interpretation and understanding of the spectral roll-off observed by the Cassini CIRS instrument (Spilker et al., 2005). Similarly, lack of laboratory data on aerosol formation in the upper atmosphere of Titan still hamper the interpretation and the understanding of the spectra collected by the Cassini INMS and CAPS instruments (Coates et al., 2007; Cui et al., 2008; Waite et al., 2005; Waite et al., 2007). The mechanism of plumes from the tiger stripes of Enceladus is controversial and unexplained (Kieffer et al., 2009; Mousis et al., 2009; Postberg et al., 2009; Schneider et al., 2009; Tyler, 2009). Laboratory studies are necessary to simulate these plumes and understand their potential origins.

The next-generation of planetary science is being planned with groundbreaking science and technology even more ambitious than in the past few decades. It is important to note, however, that any instrument on board planetary missions is a product of extensive laboratory studies, involving optimization of sensitivity, accuracy, and reproducibility. Thus, the role played by laboratory studies for planetary sciences, in both supporting the mission science and developing new instruments, is invaluable.

***It is general consent among the planetary science community that the future missions should focus more on in-situ investigations through rovers, landers, soft penetrators, and impactors, as suitable and required to obtain ground truth and subsurface composition.*** In order to achieve the next-generation in-situ science objectives in planetary sciences, a strong backing by laboratory studies is necessary. These studies not only lay down the science basis for the mission data interpretation, but they also lead to new instrument and technology development starting from laboratory scale proof of concepts at Technology Readiness Level zero (TRL0) to the most optimized and fully space qualified instrument at TRL9. Integrating the lowest TRL instrument concepts will enable a constant inflow of new

mission instruments, some of which will make to the higher TRL levels. At present we do not have enough new instruments that have been through this process, which normally takes 5 to 10 years of incubation and optimization time.

Scientists spend significant amount of time writing the proposals and reviewing them as well. When the funding agencies are constrained not to infuse sufficient funds into the programs resulting in very low funding rates (<20%), a tremendous amount of time, efforts, and resources is wasted. New ideas such as preproposals, adequate funding over a longer period of time, removing the redundant parts of the proposal writing that consumes more time, are a few suggestions that can be implemented by the funding agencies to improve the returns of tax dollars.

The Working Group on laboratory Astrophysics of the AAS (WGLA) promotes the coordination of research and knowledge between astronomy and other branches of science (physics, chemistry, geology, and biology) and is guided by advancing astronomy through laboratory astrophysics and astrochemistry to understand and interpret observational data as well as refine the models that are used to carry out simulations. Recently the WGLA has been expanded to represent the laboratory studies in support of planetary sciences. In the following we will elaborate the role played by laboratory studies and the future needs for planetary sciences and our recommendations.

## Executive Summary and Recommendations

Processes that occur at each body in our solar system can conveniently be divided into atmospheric, surface, and subsurface processes. Of course all these three regions interact with each other and their processes are interdependent and correlated. To the first approximation such a three-tier picture makes it convenient to separate the corresponding science questions and laboratory studies.

Observing, characterizing, and understanding planetary atmospheres are key components of solar system exploration. A planet's atmosphere is the interface between the surface and external energy and mass sources. Understanding how atmospheres are formed, evolve, and respond to perturbations is essential for addressing the long-range science objectives of identifying the conditions that are favorable for producing and supporting biological activity, managing the effects of human activity on the Earth's atmosphere, and planning and evaluating observations of extra-solar planets.

Due to significant work done on Earth's atmospheric processes (such as ozone dynamics, OH, O, $O_2$, halogens, CO, $CO_2$, $SO_2$, $N_2$, N, volatile organic compounds, etc.), including a large volume of laboratory work on collisional processes, atmospheric photochemistry, and general circulation models (GCMs), there is a large volume of data available on small molecules for potential direct application to atmospheres of other solar system objects. However, some (if not most) of this data are often only available at temperature and pressure ranges pertinent to Earth's atmospheric conditions which means that laboratory experiments must be done under conditions relevant to other solar system objects. More importantly, laboratory data are entirely lacking for larger, more complex, molecular aerosols and laboratory studies are currently urgently needed in this domain to support recent space mission return data.

Compared to the atmosphere, surface processes are even less understood. Most of the solar system bodies have either water ice or rock/minerals on the surface (Titan has more organic ices and a complex interaction with atmospheric organic aerosols). On those bodies with thin atmospheres and those, which are devoid of an atmosphere, radiation penetrates down to the surface. Radiation (electrons, solar wind, hard UV photons, etc.) induced processing of icy and mineral surfaces have been poorly understood and extensive laboratory studies are needed in order to obtain a comprehensive knowledge on how surfaces of solar system bodies evolve.

One area that connects surface with atmosphere, namely, gas-grain chemistry and aerosol formation, needs further laboratory work. For example, certain reactions (ex: recombination of CO and O) occur much faster through third body collisions or on the surface. These surface catalysis processes could play an important role in the exchange of material between the surface and the atmosphere.

Interiors of solar system bodies are by far least quantified, though models predict approximate constitutions. We are better off with large bodies with high densities (such as Europa, Ganymede, Mars, etc.) than small bodies with lower densities such as Enceladus. The grain-size of interior ice, minerals, and salts play an important role in convection processes. Laboratory studies with various grain-size distributions are necessary to support and refine interior models. Whether Jupiter has a solid core or not is still an open question (Saumon and Guillot, 2004). Laboratory data on the compressibility of hydrogen/deuterium at pressures pertinent to the gas giants will resolve this puzzle to a great extent. Least understood are the interiors of comets; whether the interior is made of amorphous or crystalline ice, what kind of porosity, how organics, silicates, and other minerals mix with the ice, whether the interior is uniform or not are few questions that need extensive laboratory data that could be used to refine the models and simulations to fit the observational results.

**Recommendations:**

Increase funding for laboratory studies that have specific goals to support planetary sciences, including instrument development in support of planetary missions. Specifically atmospheric, surface and subsurface relevant laboratory studies including the interface between the three phases need to be given high priority. The following laboratory studies, not ranked by any order of priority, need to be strongly supported:

Undertake studies **under simulated planetary conditions**, including:
- Laboratory studies that are related to in situ missions
- Development of spectroscopic databases covering mm waves to X-rays.
- Derivation of optical constants of cryogenic ices, organics, minerals, salts, and a mixture thereof in the 0.1 – 500 microns region.
- Detection of isotopes with high sensitivity.
- Understanding reaction mechanisms.
- Formation of aerosols from molecular precursors in planetary atmospheres.
- Better understanding of surface catalysis reactions involving gas-phase species.

- Simulations of interior processes.
- Evolution of organics on Mars and outer solar system;

Enhance current programs that support single investigators, create or strengthen Research and Analysis (R&A) programs for equipment, technology, and instrumentation development, and create Institutes of Laboratory Studies for Planetary Sciences similar to NAI (NASA Astrobiology Institute) and NLSI (NASA Lunar Science Institute). Single investigator programs are critical for university research where the next generation of laboratory astrophysicists will be trained. Adequately funded advanced technology and instrumentation programs are needed to keep pace with current observational needs and to help develop future mission capabilities. Institutes will focus on laboratory studies pertinent to specific planetary sciences objectives, such as Laboratory Studies for Titan, Laboratory Studies for Icy Satellites etc., or a NASA/NSF Consortium for Laboratory Studies (NCLS), which will have regular calls for "focused laboratory studies" pertinent to planetary sciences of strategic importance such as a forthcoming flagship mission. Under these "focused laboratory research programs" funding should also be given for building new laboratory instruments or establishing new laboratory techniques, similar to major equipment programs of NASA and NSF.

## Inner Solar System (Mercury, Venus, Mars, Earth, and the Moon)

### The Moon:

With the Moon being in the forefront again, human space explorations are planned utilizing the Moon as the space base. In order to achieve the goals of "back to the Moon" with great success, significant amount of lunar basic science and technology need to be established. Several aspects of lunar science need laboratory simulations that help future human lunar exploration. These include the ability to undertake fast in-situ surface mineral analysis and geological surveying for dust mitigation assessment, search for oxygen resources, radiation protection for human lunar explorations, subsurface water resources, lunar dust properties, etc. Finding water or the lack thereof on the Moon will have a significant impact on the scale of human lunar exploration. Thus, laboratory studies on remote-sensing subsurface water, extracting water from lunar soil (if present in the form of hydrates), and real-time analysis of lunar soil composition and in-situ chemistry that converts lunar surface and subsurface resources into energy and food would play crucial roles in future lunar human exploration.

Recommendations:

Support the following laboratory studies through increasing the size of existing funding or through new funding programs. Encourage specifically those laboratory studies that also involve proof of concept at TRL0 instrument development.

1) Fast detection and efficient extraction of water from lunar resources
2) Detection of organics on lunar surface
3) Quick identification and contrasting of lunar surface minerals
4) Properties of lunar dust – dust activation
5) Dust mitigation techniques
6) In-situ lunar resource utilization (ISRU)

**Mercury:**

Since Mercury is an airless body, surface geology is the most important laboratory experimental need. Optical properties of these minerals are urgently needed.

**Venus:**

Venus has a very dense atmosphere and is mostly viewed as an atmospheric system. Earth's atmospheric GCMs are successfully used to understand the dynamics of Venus' atmosphere. However, optical properties of oxygen containing molecules under high pressure and high temperature are necessary. Also gas-phase rate constants for certain exotic reactions that are not common in Earth's aeronomy may also need to be given priority in the laboratory.

**Recommendations:**

Laboratory studies that specifically aim at deriving rate constants or photochemical reaction pathways under the extreme conditions for Venus, not covered in the literature should be given high preference.

**Mars:**

Martian surface geology has been extensively studied by the on-site rovers Spirit and Opportunity. MRO is delivering unprecedented high-resolution images of the Martian surface. Phoenix found ice in its landing site (Smith et al., 2009). Martian atmospheric processes seem to be reasonably well understood (Terada et al., 2009; Whiteway et al., 2009; Withers, 2009). Traces of methane have been found in Martian atmosphere (Mumma et al., 2009). However, the question of organics on Mars remains an unanswered puzzle. Are there subsurface organics on Mars in the icy regions or under the mineral dust? Why are there no organics on the surface? Can we detect a marker of past organics on the Martian surface? How does radiation interact with subsurface ice on Mars? These are a few questions with astrobiological significance that need both laboratory studies and future in-situ lander/rover missions to Mars.

**Recommendations:**

Laboratory studies are needed that help determine the survival of organics under Martian surface conditions such as diurnal temperature variations and penetration of hard UV solar radiation on to the surface. Global surface radiation flux determinations on Mars will be important to model the survival rate of organics on these surfaces. Sources of methane on Mars need to be further understood. Basic laboratory research with potential in-situ instrument development even at laboratory scale (TRL0) should be strongly supported.

**The Outer Solar System**

**Jupiter and the Galilean Satellites:**

With outer planets flagship missions, the Europa Jupiter System Mission (EJSM), consisting of two coordinated missions, one from the NASA – the Europa Jupiter Orbiter (EJO), and the second from European Space Agency (ESA) – the Ganymede Jupiter Orbiter (JGO), Jupiter Magnetospheric Orbiter (JMO) of Japanese Space Agency (JAXA), on the horizon with the goal of understanding the interiors of icy worlds, it is strategically important to give enough priority for laboratory work that directly answers some of the outstanding questions regarding surface and interior ices, as well as laboratory studies that help improve the specifications of the instruments that are space-flight qualified. It is also important to conduct research in the laboratory that sets the stage for better interpretation of the mission data once it starts streaming down to Earth. This is expected to occur around 2028, approximately 18 years from now. However, we need to start a broad-based research campaign soon, as the time for incubation of experimental work as well as establishing a firm basis would need far more time than one would hope and expect.

One of the most important issues with icy worlds is the radiation processing of icy surfaces. Should there be organics – whether biotic or abiotic in their origins, what would be their chances of survival on harsh radiation surfaces such as Europa? Can organics be detected via remote sensing through traditional spectroscopic orbiter based infrared and UV-VIS instruments? OR, are their chances of survival so low that we need to do subsurface drilling? The later becomes an important issue in determining the priority and urgency of an "in-situ" payload on the EJSM spacecraft.

Due to dense atmosphere of Jupiter, it has still not been possible to get data about the surface/core of Jupiter. In fact, the very existence of a solid core is in question (Saumon and Guillot, 2004). Lack of data on the high-pressure (hundreds of GPa) compressibility of hydrogen/deuterium has made this question even more difficult to answer. Based on various compressibility ratios used in the equation of state (EOS) models, Jupiter's core can be completely liquid without the need for any solid surface. However, this question can only be answered when laboratory data for the compressibility of hydrogen/deuterium, such as (Hicks et al., 2009) are available at hundreds to thousands of GPa, and such laboratory studies should be strongly supported.

**Recommendations:**

Increase funding opportunities for laboratory work, especially those with new instrument concepts, including at the laboratory scale (TRL0). The funding agencies should encourage proof of concept instrumentation proposals that focus on laboratory research. Similarly, oversubscribed programs where funding rate is low (10 – 20%) need to be supplemented with additional funds or by other programs.

EJSM flagship mission related laboratory research would be the highlight of the coming decade. Priority should be given to laboratory research that helps define instrument parameters (such as wavelength or mass range, sensitivity, field of view etc.) or the science returns of these instruments should be given priority. These include, but are not limited to:

> - Surface VUV – Far IR (0.1 to 500 μm) properties of Europa, Ganymede, Callisto, and Io.
> - Electrical and thermal properties and their dependence on the grain-size of amorphous and crystalline ices.

- Physical properties of ices containing organic and inorganic impurities, pertinent to the Galilean satellites.
- Methods to detect and quantify subsurface oceans.
- Radiation processing through solar UV photons and Jovian magnetosphere electrons and ions of ices, salts, organics, minerals, and a combination thereof.
- Connectivity between surface, sputtering, and atmosphere for the Galilean Saellites.
- Properties of high-pressure liquid hydrogen/deuterium to model Jupiter's core.
- Gas-phase data pertinent to Jovian atmosphere and magnetosphere.

**Saturnian System:**

Cassini has been delivering exceptional data from the spectacular Saturnian system – from Saturn's polar hexagonal turbulence to the plumes of Enceladus through ring waves and dynamics. Except for Saturn and Titan, the rest of the Saturnian system surfaces are covered with water-ice. Thus, understanding the physical and chemical properties of Saturnian ices under radiation is of great importance to further our knowledge of this system. Similarly, subsurface processes that result in reasonably high temperatures causing plumes on Enceladus need to be well understood. Where does the energy come from? Tidal heating alone may not be sufficient. How is this heat transported to the surface near the South Pole – to the tiger stripes? What causes the leading/trailing albedo dichotomy of Iapetus? What is the material composition of the dark side of Iapetus? What is the composition of the rings? Can we detect non-ice molecules that cause the coloration of the rings? Could these be organics? What is the interaction between the moons and rings of Saturn? Enceladus feeds the plumes and atoms such as oxygen into the rings. Do rings also feed the Moons through accretion?

Titan is a unique case that presents a rich atmospheric composition that is seen as a good model of a prebiotic Earth, thus the importance devoted to the study of this moon. Among the main questions that arise: Why is Titan heavily fractionated with organics? What is the composition of water ice on Titan? Is solar radiation the sole energy source that drives Titan's rich atmospheric and surface chemistry? How deep the solar photons (from UV to Near IR) penetrate through Titan's atmosphere to the surface? Can longer-wavelength solar photons cause chemistry at various atmospheric depths and on the surface? How do the atmosphere and the surface of Titan interact (through aerosol formation and deposit?). How do the surface and subsurface of Titan interact (through lakes and cryovolcanism)?

**Recommendations:**

Most of the recommendations for Jovian System above are relevant to the Saturnian System as well. With outer planets flagship missions on the horizon (TSSM in standby status; Discovery and/or New Frontier missions being considered) with the goal of understanding the atmosphere and interiors of Saturnian's moons, it is important to give enough priority for laboratory work that directly answers some of the outstanding questions regarding atmospheric aerosols, surface ices, interior lakes, and, more importantly, the interaction between these three components. Laboratory studies that help improve the specifications of the instruments that are space flight qualified are also required. It is important for us to understand lower atmospheric and surface photochemistry of Titan caused by solar and cosmic radiation. It is also important to conduct research in the laboratory that sets the stage for better interpretation of the mission data once it starts streaming down to Earth. There is thus a need to start a broad-based research campaign, as the time for incubation of experimental work as well as establishing a firm basis would need far more time than one would hope and expect.

Understanding the interiors of giant planets and their satellites is still in its primitive stages. Laboratory studies that provide non-existing data, such as compressibility of hydrogen/deuterium and mixing of hydrogen and helium at very high pressures, or that improve the existing data such as the thermal conductivity of high-pressure ices vs. ice/rock mixtures, should be strongly supported.

The Rings of Saturn are still filled with lots of puzzles. Most importantly, understanding the composition of the ring particles that vary in size from a few microns to several tens of meters, as well as the interactions among these ring particles resulting in wave type motions, is of great importance to be studied under laboratory conditions as possible. To understand the radiation processing of these ices by cosmic, Saturn's magnetospheric, and solar radiation needs laboratory analog experiments. Spectroscopic data on these icy ring analogs in the 0.1 to 500 micron region to cover the Cassini UVIS, VIMS, and CIRS instruments is very important.

Increase funding opportunities for laboratory work, especially those with new instrument concepts, including at the laboratory scale (TRL0). The funding agencies should encourage proof of concept instrumentation proposals that focus on laboratory research. Similarly, oversubscribed programs where funding rate is low (10 – 20%) need to be supplemented with additional funds or by other programs.

**Small Solar System Bodies**

**Comets:**

Deep Impact on Tempel 1, though solving many puzzles, has opened-up even more questions (A'Hearn et al., 2008; A'Hearn et al., 2005; Thomas et al., 2007). In general, the interior of a comet is assumed to be highly porous, loosely bound, amorphous ice mixed with mineral grains – as formed from the protostellar dust grains, with similar composition of interstellar silicate core and ice mantle grains proposed by Greenberg (Hagen et al., 1979) several decades ago. Comets are thus expected to be the connections between solar systems and the interstellar medium and contain the history of our solar system prior to its formation. However, the properties of such loosely bound icy grains in comets are not fully understood due to the lack of appropriate laboratory data, in particular on various kinds of amorphous ices, which have different physical and chemical properties based on how they are formed and what kind of porous structures they are made off. There is no "one" well defined amorphous ice, rather there are many based on how they are formed.

With short-period comets, other issues that are still unresolved include how the comet regenerates its surface ice after sublimation during its close orbit around the Sun so that the next encounter brings enough surface material to "light-up" as the surface temperature increases to ~130 – 150 K? Is there material flow from the interior to the surface? Cryovolcanism? Rupture? What are the subsurface temperatures? What is the thermal conductivity of these loosely bound icy grains? These are a few questions that still need clear answers. Temple 1 studies indicate that this comet has high thermal inertia (Groussin et al., 2007; Groussin et al., 2009).

Comets may play a significant role in Astrobiology and origins of life. Among the most plausible working hypotheses of the origins of life is the formation of prebiotic organics and development of life during early comet impacts on Earth (Bernstein et al., 2002; Caro et al., 2002; Cockell, 2006; Deamer et al., 2002; Dworkin et al., 2004; Pierazzo and Chyba, 2002). Thus, understanding the interplay between ice and organics, radiation and thermal cycling, role of silicates and other minerals in processing these ices and organics through radiation and heating, is extremely important towards understanding the origins of Life on Earth through comet impacts. Laboratory studies are the only way that we can achieve this goal.

**Recommendations:**

The following aspects need to be addressed in the laboratory in order to understand the composition, chemistry, and dynamics of comets, especially the short-period comets.

- ➢ Laboratory studies that improve models describing comet nucleus, especially the short-period ones.
- ➢ Laboratory studies that reproduce comet coma and its composition during its lighting up on its perihelion approach.
- ➢ Role of charge generated through radiation bombardment on the surface structure of the nucleus and the gas-grain properties of the coma.
- ➢ Role of high-energy radiation emission (vacuum ultraviolet and X-ray) from comets.
- ➢ Physical properties such as thermal and electrical conductivity of amorphous water ices with/without volatile impurities such as $CO_2$, $C_2H_6$, HCN, CO, $CH_3OH$, $H_2CO$, $C_2H_2$, and $CH_4$, and mineral grains such as silicates, pyroxenes etc.
- ➢ Phase transitions between various amorphous and crystalline ices (especially hexagonal and cubic) of ice.
- ➢ Physical and chemical properties of various amorphous ices, including those derived from radiation processing of crystalline ices.
- ➢ Organics and ice chemistry in comets during radiation processing and thermal cycling.

**KBOs:**

Kuiper belt objects (KBOs) are expected to be similar in their composition to comets. Most of the compositional questions that have been discussed above under comets are pertinent to KBOs. Due to their further distance from the Sun, KBOs contain solid methane and ethane in water-ice (Barkume et al., 2008; Gibb et al., 2007; Ore et al., 2009). With ambient temperature of ~ 30 K, observation of crystalline water ice features (Marboeuf et al., 2009; Ore et al., 2009), outgassing (Levi and Podolak, 2009; Schaller and Brown, 2007), and coloration (Stern, 2009) are a few yet to be resolved issues of KBO science. Solar and cosmic radiation is expected to be the energy source for the KBO evolution. Understanding KBO properties and evolution would in turn help us to better understand comets.

**Recommendations:**

Most of the laboratory studies pertinent to Comets would also benefit KBOs. In addition, the following laboratory studies are necessary and should be strongly supported.

- ➢ Low-temperature outgassing from water ices containing CO, $N_2$, $CH_4$ and $C_2H_6$.
- ➢ Possible mechanisms to form crystalline surface ices on KBOs.
- ➢ Possible causes for the coloration of KBO surfaces.

**Asteroids:**

Asteroids originating from the asteroid belt seem to have much simpler interior/surface properties. The largest asteroid known, Ceres, is predicted to harbor surface/subsurface water (McCord and Sotin, 2005) and the DAWN mission is on its way to Vesta and Ceres (Russell et al., 2007). However, so far no direct evidence was found for surface water on these objects, but the 3.06 micron features of Vesta and Ceres are attributed to mixed organic/ice surface (Vernazza et al., 2005). Smaller asteroids are expected to be less complex in their surface/interior composition and follow the typical classification into C (carbonaceous), S (stone), and M (metallic). However, a closer encounter with an asteroid and its surface spectroscopy at far higher resolution may have more surprises in store. DAWN and Rosetta are two current missions will give us a better picture of asteroid surfaces.

**Recommendations:**

- ➢ Due to the fact that significant amount of water is expected to be on these two largest asteroids, but not necessarily positively identified yet, it is very important to undertake laboratory studies that look at mixtures of water and mineral rocks and carbonaceous chondrites under simulated solar wind conditions and identify these materials spectroscopically in the 0.1 – 20 microns region.

- ➢ It is also important to develop new instrument techniques such as standoff fluorescence/Raman spectroscopy to analyze the surface composition at very high resolution. Laboratory studies in this direction should be strongly supported.